\documentclass[12pt,english,floatfix,showkeys,superscriptaddress,aps,prd,preprint]{revtex4}
\usepackage{amsmath}
\usepackage{amssymb}
\usepackage{amsbsy}
\usepackage{amsfonts}
\usepackage{amsopn}
\usepackage{amstext}
\usepackage{graphicx}
\usepackage{amssymb}
\usepackage{amsfonts}
\usepackage{amsmath}
\usepackage{graphicx}
\usepackage{color}
\usepackage{slashed}
\usepackage{esint}
\usepackage[dvips]{epsfig}
\usepackage[dvips]{graphicx}
\usepackage{float}
\usepackage{units}
\usepackage{textcomp}
\usepackage[shortlabels]{enumitem}
\usepackage{hyperref}

\begin{document}

\title{Dymnikova-Schwinger traversable wormholes}

\author{Milko Estrada\footnote{E-mail: milko.estrada@gmail.com}}\affiliation{Facultad de Ingenier\'ia, Ciencia y Tecnolog\'ia, Universidad Bernardo O'Higgins, Santiago, Chile.}

\author{C. R. Muniz\footnote{E-mail: celio.muniz@uece.br}}\affiliation{Universidade Estadual do Cear\'a (UECE), Faculdade de Educa\c c\~ao, Ci\^encias e Letras de Iguatu, 63500-000, Iguatu, CE, Brazil.}

\begin{abstract}

In this paper, we obtain new $d$-dimensional and asymptotically flat wormhole solutions by assuming a specific form of the energy density distribution. This is addressed by considering the generalization of the so-called Dymnikova model, originally studied in the context of regular black holes. In this way, we find constraints for the involved parameters, namely, the throat radius, the scale associated to the matter distribution, and the spacetime dimension, to build those wormholes. Following, we study the properties of the obtained solutions, namely, embedding diagrams as well as Weak and Null Energy Conditions (WEC and NEC). We show that the larger the dimension, the larger the flatness of the wormhole and the more pronounced the violation of these energy conditions. We also show that the corresponding fluid behaves as phantom-like for $d \geq 4$ in the neighborhood of the wormhole throat. In addition, we specialize the employed model for $d=4$ spacetime, associating it with the gravitational analog of the Schwinger effect in a vacuum and correcting the model by introducing a minimal length via Generalized Uncertainty Principle (GUP). Thus, we obtain a novel traversable and asymptotically flat wormhole solution by considering that the minimal length is very tiny. The associated embedding diagram shows us that the presence of this fundamental quantity increases the slope of the wormhole towards its throat compared with the case without it. That correction also attenuates the WEC (and NEC) violations nearby the throat, with the fluid ceasing to be a phantom-type at the Planck scale, unlike the case without the minimal length.

\end{abstract}

%\pacs{72.80.Le, 72.15.Nj, 11.30.Rd}

\maketitle
%%%%%%%%%%%%%%%%%%%%%%%%%%%%%%%%%%%%%%%%%%%%%%%%%%%%%%%%%%%%%%%%%%%%%%%%%%%%%%%
\section{Introduction}

Wormholes are hypothetical objects with non-trivial geometry and topology predicted by general relativity (GR), representing a kind of tunnel in the spacetime that connects two remote regions of the same universe or two different universes (see \cite{Morris:1988cz,Visser}, and references therein). These theoretical structures have been recently investigated in a fundamental level from the J. Maldacena works \cite{Maldacena:2013xja,Maldacena:2017axo,Maldacena:2020sxe} and also in more applied contexts of condensed matter systems \cite{Gonzalez:2009je,Alencar:2021ejd}. Usually, it is required some type of exotic matter sourcing traversable wormholes. However, in scenarios of modified theories of gravity, such a feature can change with non-exotic matter working as a source for the wormhole geometry  \cite{Pavlovic:2014gba,Myrzakulov:2015kda,Mehdizadeh:2019qvc,Sahoo:2020sva,Moti,Alencar,Sadeghi:2022sto,Nilton:2022cho,Harko:2013yb,Bronnikov:2016xvj,Kuhfittig:2020zmp,Chanda:2021dvc}. Furthermore, it is worth mentioning that, because there are several branches of theoretical physics that have predicted the existence of extra dimensions (as for example the string theory), the study of higher dimensional wormholes also has been of physical interest in the last years. See \cite{Paul:2021lvb,Oliveira:2021ypz}, for instance.

An example of a source of exotic matter that can sustain a traversable wormhole is the Casimir vacuum energy. Thus, wormholes with Casimir-type energy density profiles have been studied in the last years \cite{Garattini:2019ivd,Jusufi:2020rpw,Tripathy:2020ehi,Carvalho:2021ajy,Oliveira:2021ypz}. In particular, \cite{Jusufi:2020rpw} considered models which introduce a fundamental minimal length via Generalized Uncertainty Principle (GUP) in order to correct that energy and study wormholes solutions in the GR context.

On the other hand, finding analytic wormhole solutions to the Einstein field equations, with the presence of matter sources in the energy-momentum tensor, is not an easy task because of the highly nonlinear behavior of the equations of motion. One very common strategy consists in using different state equations in order to obtain those solutions. Some examples can be found in  references \cite{Cataldo:2015vra,Tello-Ortiz:2021kxg}. Among other examples of wormhole solutions with energy density profiles we have the Class one approach \cite{Tello-Ortiz:2020zfs}, the dark matter energy profile \cite{Xu:2020wfm,Muniz:2022eex}, Yukawa–Casimir wormholes \cite{Garattini:2021kca}, etc.

In this work, we will study new traversable and asymptotically flat wormhole solutions with local sources of matter. Thus, we will employ the $d$-dimensional generalization \cite{Estrada:2019qsu} of the Dymnikova energy density \cite{Dymnikova:1992ux}. This model has been used to generate regular black hole solutions (RBH), such that its energy is quasi-localized at infinity. Such a model of RBH is restricted to the case where $\rho=-p_r$ and $g_{tt}=g_{rr}^{-1}$. It is worth mentioning that, clearly, both the geometric and the physical characteristics of the wormhole solutions differ from the RBH solutions. In the wormhole context, the concept of quasi--localized energy at infinity seems to have no physical meaning. In this sense, our wormhole model will differ from the Dymnikova RBH model.

Therefore, we will obtain a novel family of $d$-dimensional and asymptotically flat wormhole solutions, showing that the employed material source is capable of forming traversable wormholes under specific conditions (principally the flaring-out one). We will study their principal properties, such as embedding diagrams, Weak and Null Energy Conditions (WEC and NEC), and the position-dependent state parameter. Following, we will specialize this analysis for a 4-dimensional universe and, by formally associating the Dyminikova density profile with the Schwinger particle-antiparticle pair production in a vacuum according to \cite{Dymnikova:1996plb,Ansoldi:2008jw}, we will correct the model introducing a fundamental minimal length via Generalized Uncertainty Principle (GUP) following \cite{Haouat:2013yba,Ong:2020tvo}. With this, we find a novel traversable and asymptotically flat wormhole solution and also study those properties, remarking on the role of the minimal length in them.

The paper is structured as follows: In  Section II we review the basic features of $d$-dimensional traversable wormholes and their sources. In section III we enunciate a list of conditions that must satisfy our model. In Section IV, we present our model and build new $d$-dimensional wormhole solutions, studying the corresponding properties. In Section V we present and study Dyminikova-Schwinger GUP-corrected wormhole solutions. Finally, in Section VI, we conclude the paper.

\section{A short review of $d$-dimensional wormholes}

The following line element describes the geometry of a static and spherically symmetric Lorentzian traversable wormhole in a spacetime $d$-dimensional \cite{Oliveira:2021ypz}:
\begin{equation}
ds^2=-e^{2\Phi(r)}dt^2+\frac{dr^2}{1-b(r)/r}+r^2d\Omega^2_{d-2},
    \label{metric}
\end{equation}
where $\Phi(r)$ and $b(r)$ are arbitrary functions of the radial coordinate, $r$, denoted as the redshift function, and the shape function, respectively. That coordinate decreases from infinity to a minimum value $r_0$, the radius of the throat, where $b(r_0)=r_0$. The quantity $d\Omega_{d-2}$ is the solid angle element in $d-2$ spacetime dimensions. 

For this spacetime to represent a wormhole solution, the following conditions must be satisfied\cite{Cataldo:2015vra}
\begin{enumerate}
    \item \label{restriccion1} The function $\Phi(r)$ must be finite for all value of $r \ge r_0$ in order to avoid singularities and horizons.
    \item  \label{restriccion2} The minimum value of $r$ at the throat of the wormhole, corresponds to the point $r=r_0=b(r_0)$, where the function $g_{rr}^{-1}=1-b(r)/r$ vanishes
    \item \label{restriccion3} The  proper radial distance
    \begin{equation}
        l(r) = \pm \int_{r_0}^r \frac{dr}{\sqrt{1-\frac{b(r)}{r}}}
    \end{equation}
    must be finite. For this it is necessary that
    \begin{equation}
           1-\frac{b(r)}{r} \ge 0
    \end{equation}
    for all value of $r \ge r_0$.
\item \label{restriccion4} For the asymptotic flatness condition it must be satisfied that
\begin{equation}
    \displaystyle \lim_{r \to \infty} \frac{b(r)}{r} \to 0
\end{equation}
This ensures that at infinity the proper radial distance $l \to \pm \infty$.
\end{enumerate}

\section{Our model} \label{nuestrasCondiciones}

Now we provide a list of constraints that the energy density as well as both the redshift and shape functions must satisfy for our model to represent a $d$-dimensional and asymptotically flat wormhole solution:

\begin{enumerate}[(a)]
\item \label{CondicionA} The energy density must be a continuous, positive and decreasing function such that
\begin{equation} \label{densidadinfinito}
    \displaystyle \lim_{r \to \infty} \rho(r) \to 0
\end{equation}
Furthermore, as we will see below, through the equations of motion, this energy density leads to a function $b(r)$ of the form
\begin{equation} \label{funcionb}
    b(r) = \frac{\bar{b}(r)}{r^{d-4}}
\end{equation}
where $\bar{b}(r)$ must be an increasing function, such that, due to the condition \eqref{densidadinfinito}, it must be satisfied that:
\begin{equation} \label{binfinito}
\displaystyle \lim_{r \to \infty} \bar{b}(r) = \mbox{Constant}
\end{equation}
This latter ensures condition \ref{restriccion4} of introduction.
\item \label{CondicionB} Relating with condition \ref{restriccion2} above, the largest solution of the equation $g_{rr}^{-1}=1-b/r=0$ corresponds to the minimum value of $r=r_0$ at the wormhole throat.
\item \label{CondicionC} The function $b(r)$, equation \eqref{funcionb}, must be such that the function $g_{rr}^{-1}=1-b/r$ be an increasing function from $g_{rr}^{-1}(r=r_0)=0$ up to $g_{rr}^{-1}(r \to \infty) \to 1$ (value provided by equation \ref{binfinito}). So, it must be satisfied that
\begin{equation} \label{gprimar0}
(g_{rr}^{-1})'|_{r \ge r_0} > 0
\end{equation}
This item ensures condition \ref{restriccion3} of introduction.
\item \label{CondicionD} In order to avoid singularities, with a finite value of $e^{2\Phi(r)}$, condition \ref{restriccion1} of introduction,  and in order to have an asymptotically flat behavior, we impose that
\begin{equation} \label{phir0}
 e^{2\Phi(r_0)}  =A
\end{equation}
with $0<A \le 1$, and
\begin{equation} \label{phiInfinito}
    \displaystyle \lim_{r \to \infty} e^{2\Phi(r)} =1
\end{equation}
Furthermore the derivative
\begin{equation} \label{phiPrima}
\left (e^{2\Phi(r)} \right)' \ge 0
\end{equation}
for $r\ge r_0$. This ensures that the function $e^{2\Phi(r)}$ can be either a constant (for $A=1$ and $\left (e^{2\Phi(r)} \right)' =0$, {\it i.e} the zero tidal case) or an increasing function. Furthermore we impose that :
\begin{equation} \label{phiPrimaInfinito}
    \displaystyle \lim_{r \to \infty} \Phi' =0
\end{equation}
\end{enumerate}

%\subsection{The generic solution}

For the line element \eqref{metric}, the $(t,t)$ component of the $d-$ dimensional field equations is:
\begin{equation}\label{00-EE}
\left (r^{d-4}b(r) \right )'= \frac{2}{d-2} r^{d-2} \rho(r)
\end{equation}
where we have made Einstein's constant $\kappa=1$ and where $\rho$ is constructed such that the conditions \ref{CondicionA} of section \ref{nuestrasCondiciones} must be satisfied. This energy density  gives rise to the following solution
\begin{equation} \label{bgenerico}
    b(r)= \frac{\bar{b}(r)}{r^{d-4}}
\end{equation}
where $b$ and $\bar{b}$ must also satisfy the conditions \ref{CondicionA} of section \ref{nuestrasCondiciones} and where
\begin{equation} \label{bbarragenerico}
    \bar{b}(r)=\frac{2}{d-2} \int r^{d-2} \rho (r) dr
\end{equation}

Using $b(r)$ and $\Phi(r)$ (which are constructed such that the conditions \ref{CondicionD} must be satisfied), we can determinate the radial pressure using the $(r,r)$ component
\begin{equation} \label{presionradialGenerico}
p_r = \frac{d-2}{2r^2} \left [ \left ( 1- \frac{b}{r}  \right) \left ( 2 r \Phi' + (d-3)    \right ) - (d-3) \right ]
\end{equation}
Furthermore, from equations \eqref{binfinito} and \eqref{phiPrimaInfinito} :
\begin{equation} \label{prinfinito}
\displaystyle \lim_{r \to \infty} p_r \to 0
\end{equation}

The energy-momentum tensor has the form $T^\mu_\nu = \mbox{diag} (-{\rho}, p_r, p_\theta, p_\phi, ...)$. From spherical symmetry we have for all the $(d-2)$ angular coordinates $p_t=p_\theta=p_\phi=...$ and, the conservation law $T^{AB}_{;B}=0$ gives:
\begin{equation}
( \rho + p_r) \Phi' + p'_r+\frac{d-2}{r}(p_r-p_t ) =0 . \label{conservacion1}
\end{equation}
So, using the last equation can be determined the tangential pressure values. It is direct to check that
\begin{equation} \label{ptinfinito}
    \displaystyle \lim_{r \to \infty} p_t =0
\end{equation}

It is worth mentioning that, these constraints should serve to construct several types of higher dimensional and asymptotically flat wormhole solutions. Bellow, as a particular example we will use the $d$-dimensional generalization \cite{Estrada:2019qsu} of the Dymnikova energy density \cite{Dymnikova:1992ux}. However there are several examples of energy density models that comply the constraints described in this section (for example \cite{Aros:2019quj,Spallucci:2017aod}). Furthermore, a particular form for the redshift function will be proposed. However, following these constraints, other models of energy density as well as other forms for the redshift function, could serve for construct new asymptotically flat wormhole solutions in future works.

\section{The New family of wormhole solutions}

So far it has been shown a suitable model to describe an higher dimensional and asymptotically flat wormhole solution. Thus, we have not proposed a specific form for the energy density $\rho$ and the $\Phi$ function, and for the further values of $b(r)$ , $p_r$ and $p_t$. Thus, we will employ here the $d$-dimensional generalization \cite{Estrada:2019qsu} of the Dymnikova energy density \cite{Dymnikova:1992ux}. In four dimensions this model of energy density has give rise to regular black hole solutions, as it was already mentioned in Introduction, as well as to wormhole magnetic monopoles,  under the constraints $\rho=-p_r$, studied in reference \cite{Romero:2019ull} in a different framework from what is made here.

The energy density profile is given by
\begin{equation} 
\rho(r)=\frac{d-2}{2}\rho_0\exp{\left(-\frac{r^{d-1}}{a^{d-1}}\right)} \label{DensidadDymnikova1}
\end{equation}
where $\rho_0,a>0$ are constants. It is direct to check that this model of energy density is consistent with the conditions \ref{CondicionA}.

Substituting in equation \eqref{bbarragenerico}:
\begin{equation}
\bar{b}(r)= \frac{a^{d-1}\rho_0}{(d-1)}\left[1-\exp{\left(-\frac{r^{d-1}}{a^{d-1}}\right)}\right].
\end{equation}
where has been used the value $\frac{a^{d-1}\rho_0}{(d-1)}$ as integration constant. It is direct to check that condition \eqref{binfinito} is satisfied.

Thus, it is worth mentioning that the conditions \ref{CondicionA} of section \ref{nuestrasCondiciones} are satisfied.

From equation \eqref{bgenerico}
\begin{equation} \label{bDymnikova1}
b(r)=\frac{a^{d-1}\rho_0}{(d-1)r^{d-4}}\left[1-\exp{\left(-\frac{r^{d-1}}{a^{d-1}}\right)}\right].
\end{equation}

In order to satisfy the condition \ref{CondicionB} of section \ref{nuestrasCondiciones},  one has $b(r_0)=r_0$, where $r_0$ is the throat radius, so we must find a relationship between $\rho_0$, $d$, $a$, and $r_0$, which is given by
\begin{equation} \label{rho0Dymnikova}
\rho_0=\frac{(d-1)a^{1-d}r_0^{d-3}}{1-\exp{\left(-\frac{r_0^{d-1}}{a^{d-1}}\right) }}.
\end{equation}

Thus, the $g_{rr}^{-1}$ metric component is
\begin{equation}
    g_{rr}^{-1} =   1-\frac{r_0^{d-3}}{r^{d-3}} \left[ \dfrac{ 1-\exp{\left(-\frac{r^{d-1}}{a^{d-1}}\right)} }{   1-\exp{\left(-\frac{r_0^{d-1}}{a^{d-1}}\right) }         }   \right]
\end{equation}

\begin{figure}[ht!]
\centering
 	\includegraphics[width=0.5\textwidth]{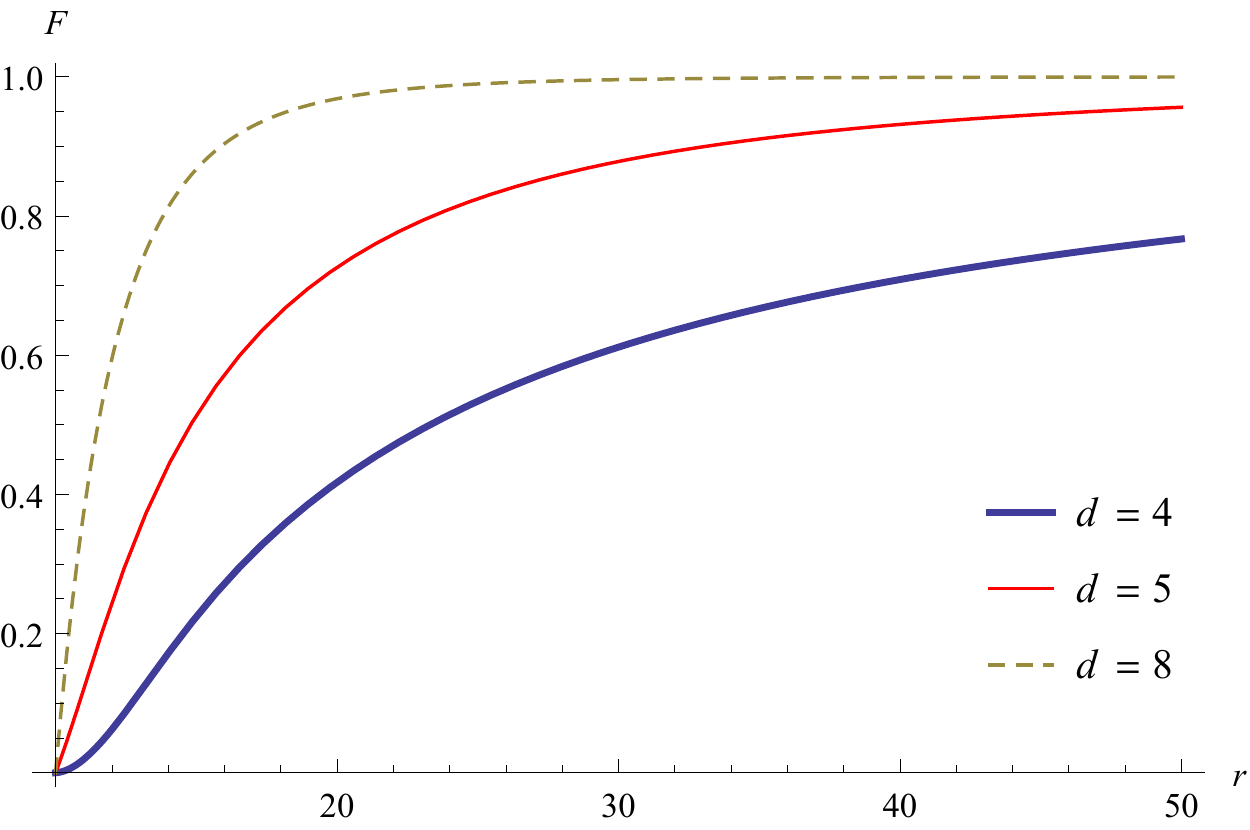}
 	\caption{Behavior of the function $F(r)=1- b(r)/r$, for some spacetime dimensions. The parameter set is $r_0=10.0$ and $a=8.0$.}
 	\label{Cond1}
 \end{figure}

As it was mentioned in conditions \ref{CondicionB} and \ref{CondicionC}, $r_0$ must represent the largest solution of the equation $g_{rr}^{-1}=0$ and furthermore the derivative $(g_{rr}^{-1})'|_{r=r_0}$ must be positive. It is easy to check that, for $r \in [0, \infty]$, there is a critical  value $r=r^*$, where $r^*$ represents a point where the function $g_{rr}^{-1}$ reaches a  local minimum. So, $r_0$, such that $g_{rr}^{-1}(r=r_0)=0$,  must be located after the critical value, {\it i.e} $r_0>r^*$. This occurs provided $r_0/a > \beta(d)$, where $\beta(d)$ is a number of order of unit and that can be calculated numerically from the Lambert function, $W_{-1}(z)$. As for example for $d=4$, $\beta(4)\approx 1.24$, for $d=5$, $\beta(5)\approx 1.06$, and for $d=8$, $\beta(8)=0.94$. So, the conditions \ref{CondicionB} and \ref{CondicionC} are satisfied. Fig. \ref{Cond1} shows us the behavior of the function $g_{rr}^{-1}=F(r)=1-b(r)/r$. Notice that the obtained wormhole solutions are asymptotically flat, since when $r\to\infty$, $b(r)/r\to 0$, for all $d$. 

As redshift function $\Phi(r)$ we choose in arbitrary way the following form:
\begin{equation} \label{funcione2Phi}
    e^{2\Phi}(r)=1-C \exp \left (-\frac{r-r_0}{a} \right )
\end{equation}
where it is easy to check that $C=0$ represents the zero-tidal case. It is easily noted that the condition \eqref{phiInfinito} is satisfied.
The function evaluated at $r=r_0$ is:
\begin{equation}
    e^{2\Phi}(r=r_0)=1-C=A
\end{equation}
which in order to satisfy \eqref{phir0} must satisfy that
\begin{equation}
    0 \le C < 1
\end{equation}
Furthermore it is direct to check that the conditions \eqref{phiPrima} and \eqref{phiPrimaInfinito} are satisfied. So, this function serves as test of prove in order to satisfy the condition \ref{CondicionD}.

The value of the radial pressure is obtained easily by substituting equations \eqref{rho0Dymnikova}, \ref{bDymnikova1} and \eqref{funcione2Phi} into the equation \eqref{presionradialGenerico}. The tangential pressure is computed from equation \eqref{conservacion1}.

\subsection{Flaring-out condition}
A fundamental property of a wormhole is that a flaring-out condition of the throat, given by $(b-b^{\prime}r)/b^{2}>0$ \cite{Morris:1988cz}, and at the throat
$b(r_{0})=r=r_{0}$, such that the condition $b^{\prime}(r_{0})<1$ is imposed to have wormhole solutions. It is precisely these restrictions that impose the WEC and NEC violations in classical general relativity. For our model, using the condition \eqref{gprimar0}, {\it i.e.} $(1-b/r)'|_{r=r_0}>0$ it is direct to check that the condition $b^{\prime}(r_{0})<1$ is automatically satisfied, and thus, the flaring out condition is satisfied for $r=r_0$ . Furthermore, using the same condition for $(1-b/r)'|_{r>r_0}>0$ it is also direct to check that the flaring out condition is satisfied for $r>r_0$.

\subsection{Embedding diagram}

\begin{figure}[!ht]
    \centering
    \begin{minipage}{0.5\linewidth}
        \centering
        \includegraphics[width=1.2\textwidth]{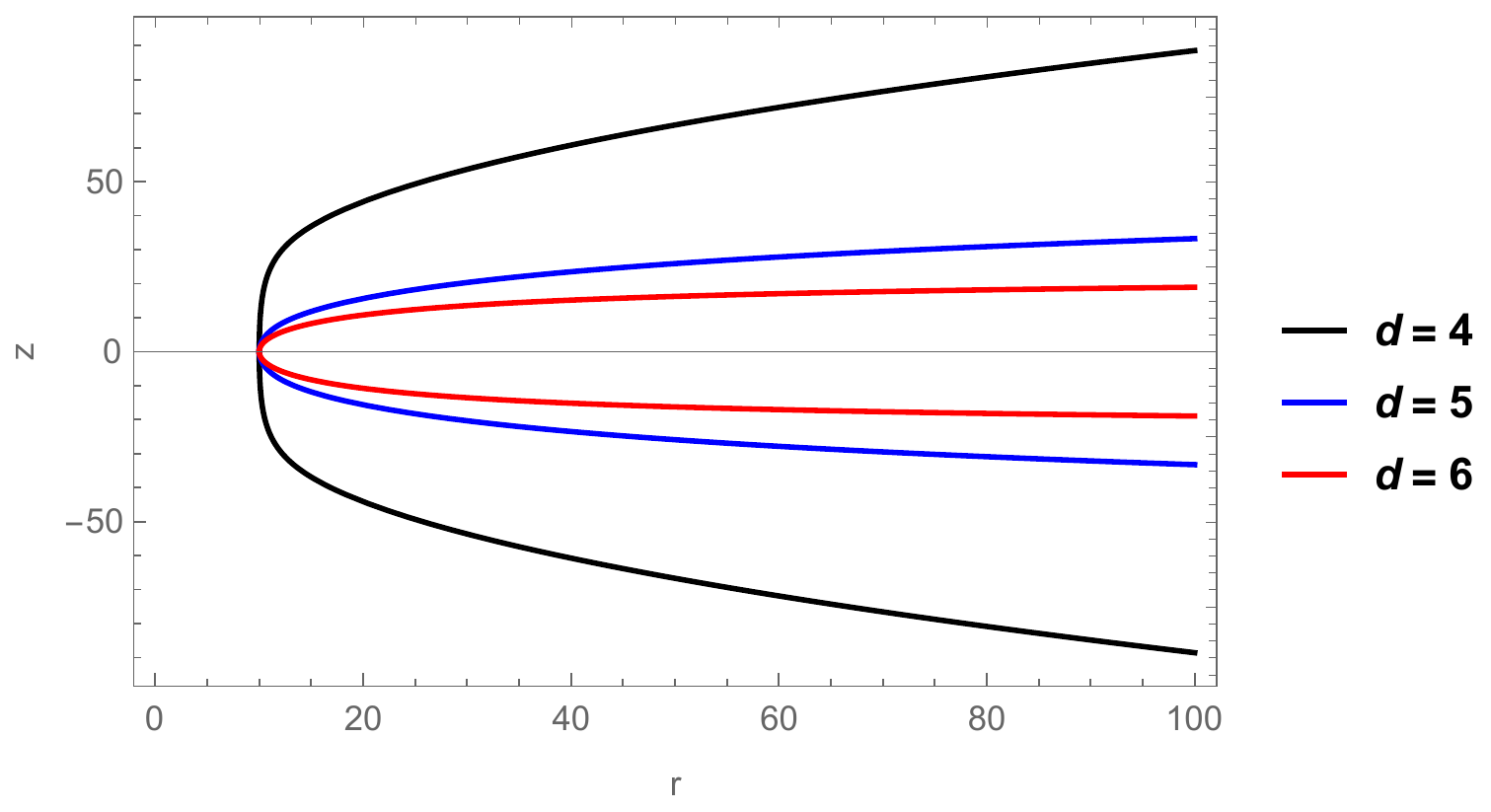}
        \label{fig:figura1minipg}
    \end{minipage}\hfill
   \caption{Embedding diagram. The parameter settings are $a=8$ and $r_0=10$, in Planck units for $d=4,5,6$.}
    \label{embedding1}
\end{figure}

This diagram is usually obtained by comparing the spatial three-dimensional flat metric written in cylindrical coordinates ($r,\phi,z)$ with the spatial sector of the Morris-Thorne metric (fixing the polar angle at $\theta=\pi/2$). The same reasoning can be generalized for a $d$-dimensional spacetime, with the flat spatial sector being compared with the $d-1$-dimensional spatial sector of the spherical wormhole (fixing also the $(d-3)$ polar angular coordinates at $\pi/2$), and identifying the azimuthal angles, arriving at
\begin{equation}\label{z-r}
dz=\pm dr\sqrt{\frac{b(r)/r}{1-b(r)/r}}.
\end{equation}.
By means of equation \eqref{binfinito} it is easy to check that
\begin{equation}
\displaystyle \lim_{r \to \infty}    \frac{dz}{dr}= 0
\end{equation}
We can see the generic behavior in figure \ref{embedding1}. We can note that in higher dimensions the flatness of the wormhole is bigger.

\subsection{Energy conditions and linear parameter state}
\begin{figure}[!h]
    \centering
    \begin{minipage}{0.5\linewidth}
        \centering
        \includegraphics[width=1.0\textwidth]{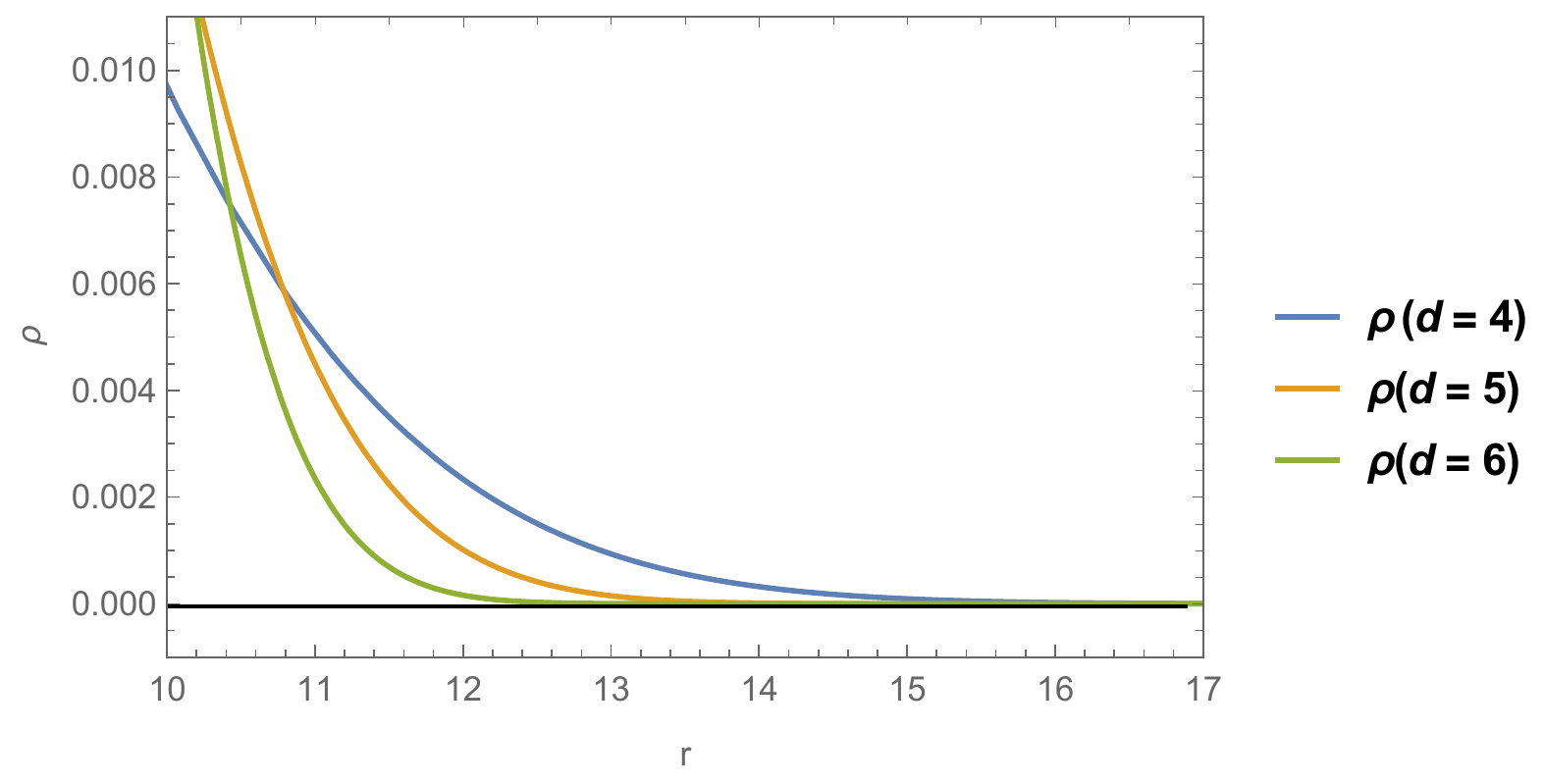}
        \label{Rho}
    \end{minipage}\hfill
    \begin{minipage}{0.5\linewidth}
        \centering
        \includegraphics[width=1.1\textwidth]{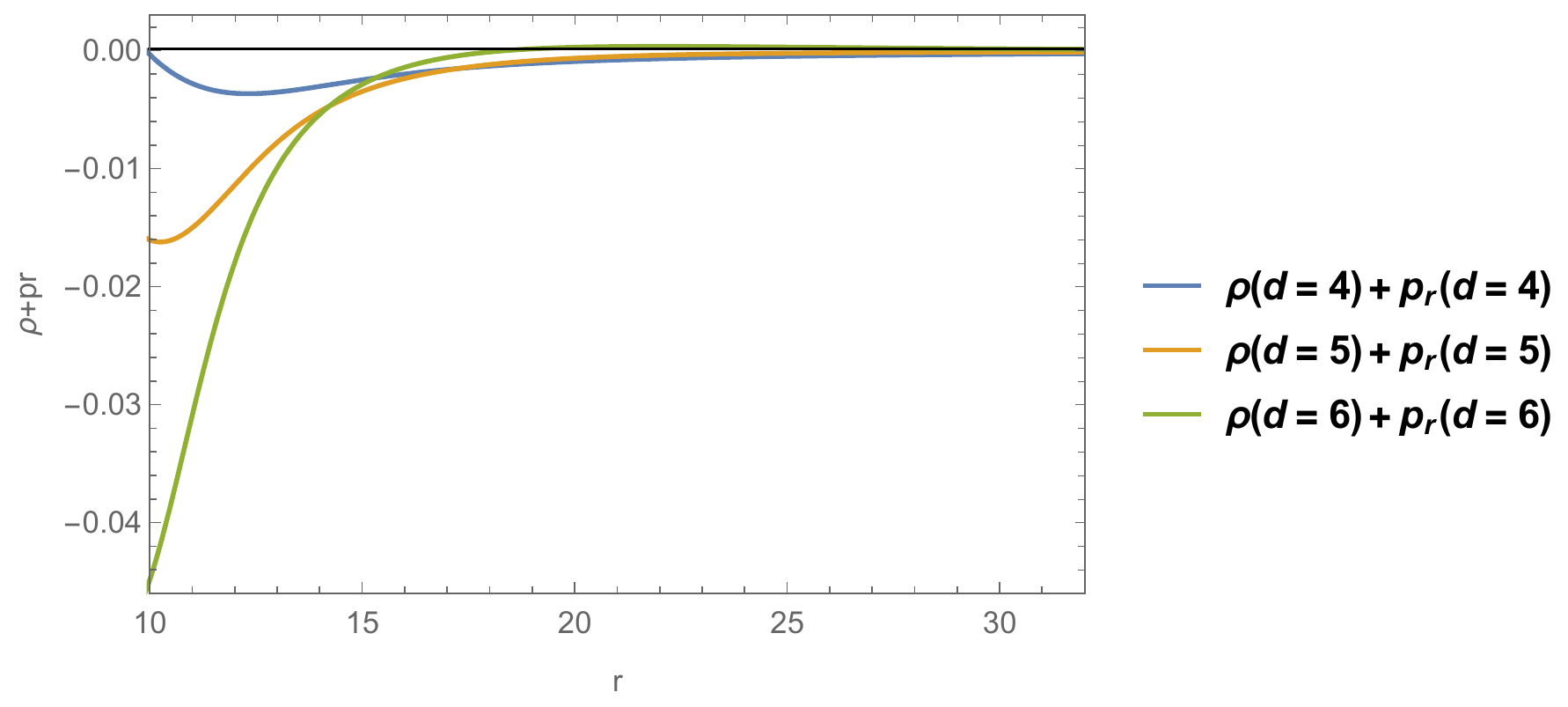}
              \label{Rho+pr}
    \end{minipage}
     \begin{minipage}{0.55\linewidth}
        \centering
        \includegraphics[width=1.2\textwidth]{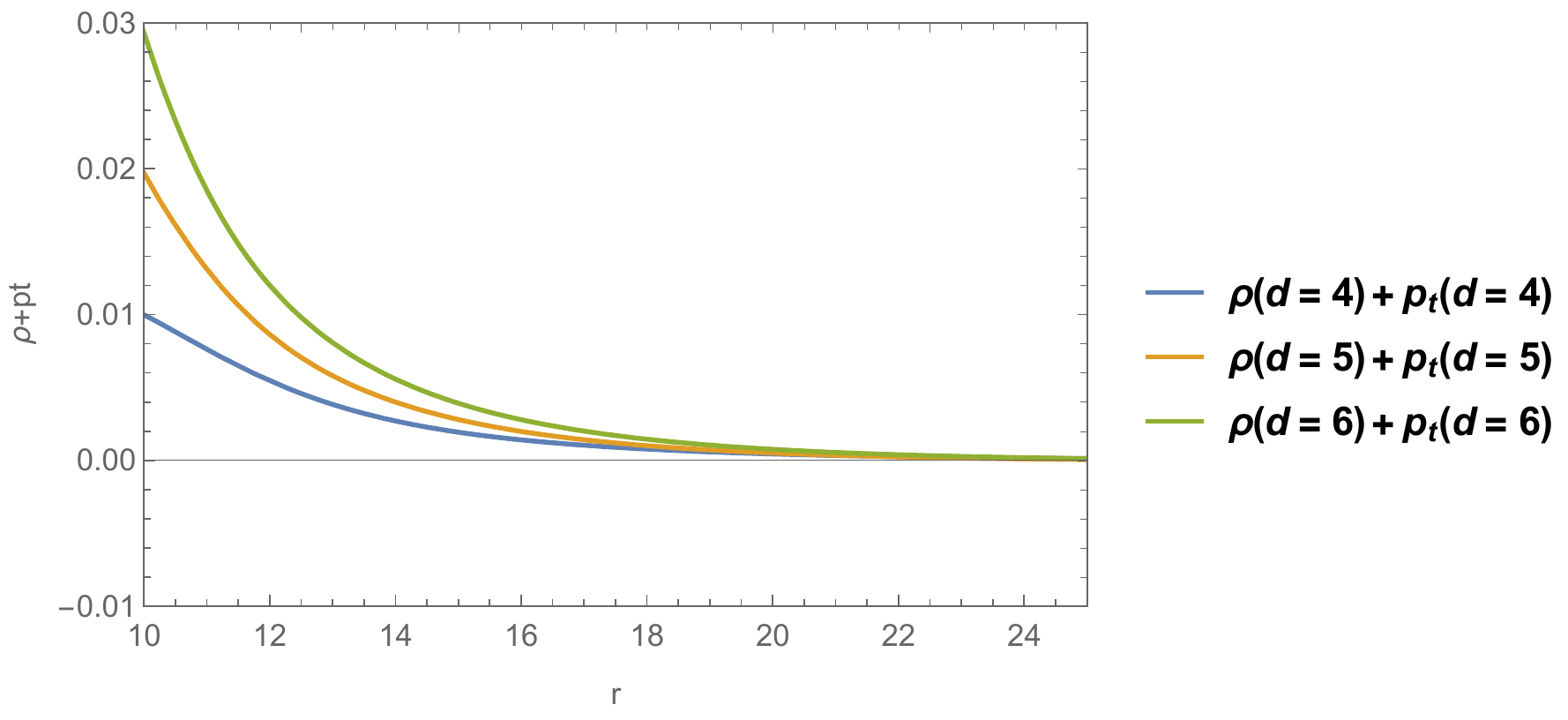}
              \label{Rho+pt}
    \end{minipage}
      \caption{Fist figure : $\rho$. Second figure : $\rho+p_r$. Third figure : $\rho+p_t$. The parameter set is $r_0=10.0$, $a=8.0$, $C=0.6$ and $d=4,5,6$. }
    \label{EnergyD}
\end{figure}
In figures \ref{EnergyD} we can test the behavior of the Null Energy Conditions (NEC $\rho+p_i \ge 0$) and Weak Energy Conditions (WEC $\rho\ge 0,\rho+p_i \ge 0$). We can notice that such behaviors are generic for other values of $d$. In the second figure we can note that WEC (and NEC) are violated because the condition $\rho+p_r \ge 0$ is only satisfied for $r \to \infty$. 

We can also analyze the exoticity of the source by considering the state parameter of a perfect fluid obeying the linear EoS $\omega(r)=p_r/\rho$, in the neighborhood of the wormhole throat. Thus, we have that \begin{equation}
\lim_{r\to r_0}\omega(r)=\frac{3-d}{d-1} \cdot \frac{\exp (r_0/a)^{d-1} -1}{(r_0/a)^{d-1}} . 
\end{equation}
It is direct to check that, following the condition $r_0/a > \beta(d)$, described above, the value of $\omega(r)<-1$. For instance, in $d=4$ the source behaves like a phantom fluid since the flaring-out condition predicts that $r_0 \gtrsim 1.24 a$, as mentioned more above, and then we can show that $\omega(r)<-1$. Below, we will see that, still in the $4d$ scenario, the presence of the minimal length introduced by means of the Generalized Uncertainty Principle (GUP) modifies this behavior near the throat.

\section{Dymnikova-Schwinger 4D wormhole and the influence of a minimal length}

According to \cite{Dymnikova:1996plb,Ansoldi:2008jw}, the $d=4$ Dymnikova density profile can be seen as the gravitational analogue of the electron-positron pair production rate, $\Gamma\sim \exp{(-E_c/E)}$, in the vacuum -- the so-called Schwinger effect. This QED phenomenon is associated with the application of an intense uniform electric field ($E$) that results in vacuum polarization and the corresponding production of particle pairs. The critical electric field necessary for an abundant pair production is given by $E_c=\pi\hbar m_e^2/e$, where $m_e$ and $e$ are the electron mass and charge, respectively. The gravitational equivalent is considered when one makes the association of the electric field with the gravity tension characterized by a curvature term, namely $E\sim r^{-3}$ and $E_c\sim a^{-3}$, so that we obtain the $d=4$ Dymnikova-Schwinger density profile of Eq. (\ref{DensidadDymnikova1}).

The correction to the Schwinger effect associated to the existence of a minimal length was obtained in \cite{Haouat:2013yba,Ong:2020tvo} by means of the Generalized Uncertainty Principle (GUP). In this case, the electron-positron pair production rate becomes
\begin{equation}
\Gamma\sim \exp{\left(-\frac{A}{E}+B(\alpha) E\right)},
\end{equation}
where $A$, $B(\alpha)$ are constants depending on the mass and charge of the electron, $E$ is the electric field and $\alpha$ comes from GUP via
\begin{equation}
\Delta x\Delta p\sim \frac{\hbar}{2}\left[1+\frac{\alpha(\Delta p)^2}{\hbar}\right],
\end{equation}
with $\alpha=\ell^2$. Here $\ell$ is the minimal length. Identifying the electrical field with the gravitational tension as it was previously discussed, the GUP correction to the Dymnikova-Schwinger density profile will be, therefore
\begin{equation}\label{GUPCorr}
\rho(r)=\rho_0 \exp{\left(-\frac{r^3}{a^3}+\frac{\alpha a}{r^3}\right)}\approx \rho_0\exp{\left(-\frac{r^3}{a^3}\right)}\left(1+\frac{\alpha a}{r^3}\right),
\end{equation}
where we have again considered the scale of the matter distribution, $a$, and the minimal length via $\alpha$. In the second relation, we taken into account that this minimal length is very tiny, $\alpha/r^2\ll 1$ (recalling that $r\geq r_0$).

The shape function of Dymnikova-Schwinger GUP-corrected wormhole can be obtained from the $(t,t)$ component of Einstein's equations, (\ref{00-EE}), for $d=4$, with the energy density profile (\ref{GUPCorr}), and it can be written as
\begin{equation}\label{b(r)2}
b(r)=r_0\frac{\left[1-\exp{\left(-\frac{r^3}{a^3}\right)}+\frac{\alpha}{a^2}Ei\left(-\frac{r^3}{a^3}\right) \right]}{\left[1-\exp{\left(-\frac{r_0^3}{a^3}\right)}+\frac{\alpha}{a^2}Ei\left(-\frac{r_0^3}{a^3}\right) \right]},
\end{equation}
so that the integration constant was chosen in order to do $b(r_0)=r_0$, with $Ei(z)$ being the exponential integral function. The metric of the wormhole under consideration is, therefore,
\begin{equation}\label{Sol2}
ds^2=-\left[1-C\exp{\left(-\frac{r-r_0}{a}\right)}\right]dt^2+\frac{dr^2}{1-\frac{r_0}{r}\frac{\left[1-\exp{\left(-\frac{r^3}{a^3}\right)}+\frac{\alpha}{a^2}Ei\left(-\frac{r^3}{a^3}\right) \right]}{\left[1-\exp{\left(-\frac{r_0^3}{a^3}\right)}+\frac{\alpha}{a^2}Ei\left(-\frac{r_0^3}{a^3}\right) \right]}}+r^2d\Omega^2,
\end{equation}
where we have taken into account the redshift function given in Eq. (\ref{funcione2Phi}).

\begin{figure}[!ht]
    \centering
    \begin{minipage}{0.5\linewidth}
        \centering
        \includegraphics[width=0.95\textwidth]{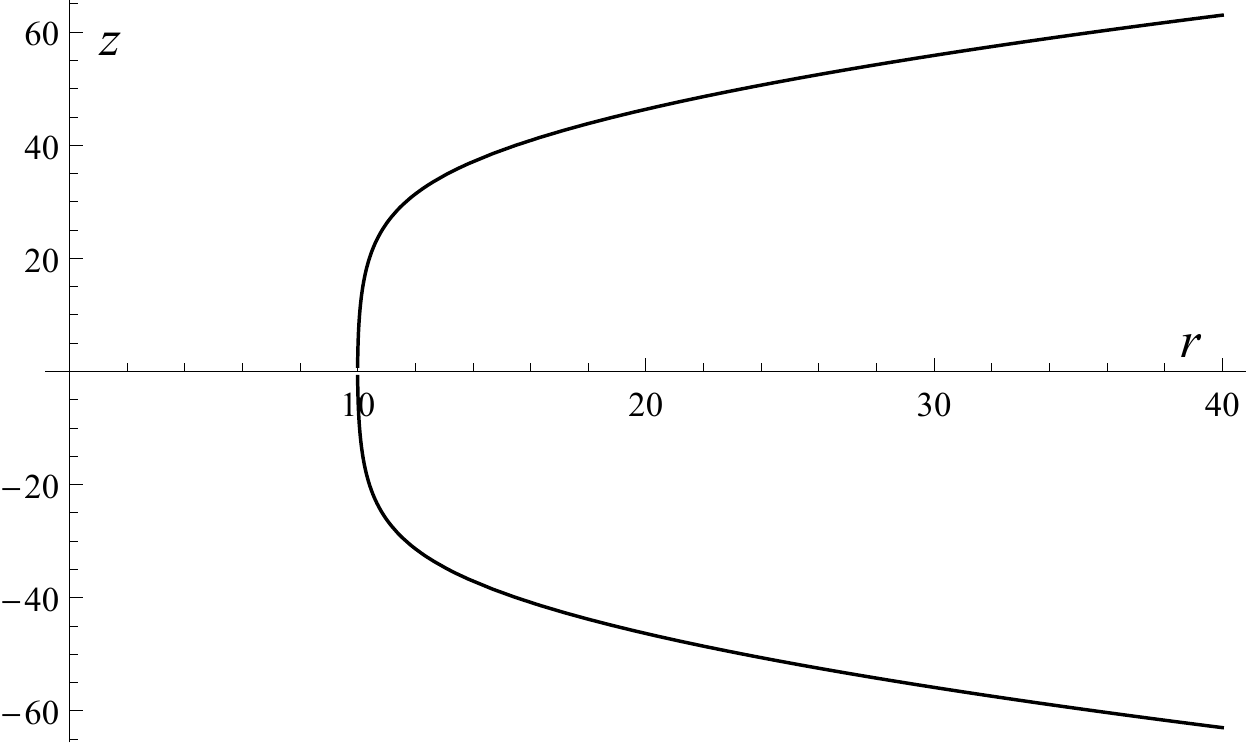}
        \label{fig:figura1minipg}
    \end{minipage}\hfill
    \begin{minipage}{0.5\linewidth}
        \centering
        \includegraphics[width=0.95\textwidth]{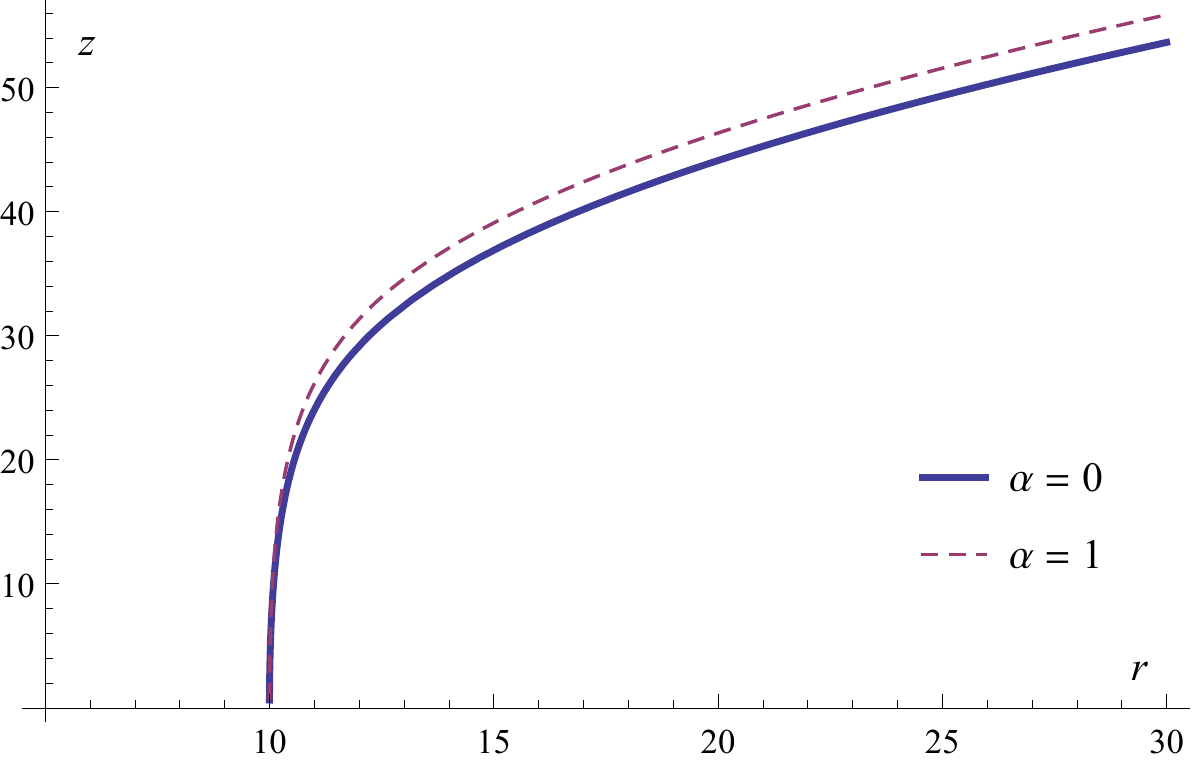}
              \label{fig:figura2minipg}
    \end{minipage}
   \caption{Embedding diagram of the Dymnikova-Schwinger GUP-corrected wormhole profile in the left panel, whose top mouth is zoomed in the right panel. This latter exhibits the greater slope of that wormhole ($\alpha\neq 0$) when compared to the one without a minimum length ($\alpha=0$).  The parameter settings are $a=8.0$ and $r_0=10.0$, in Planck units.}
    \label{embedding}
\end{figure}
In Fig. \ref{embedding} we depict the embedding diagram profiles of the Dymnikova-Schwinger GUP-corrected wormhole. 
%Substituting into the above equation the expression (\ref{b(r)2}) and integrating it, we obtain the corresponding embedding diagrams, for both $\alpha=0$ and $\alpha\neq 0$ Dymnikova-Schwinger wormholes. By considering these diagrams, we note that the presence of the minimal length increases the slope of the wormhole, which is more accentuated the greater this length.

\subsection{Energy conditions and linear state parameter}
\begin{figure}[!h]
    \centering
    \begin{minipage}{0.5\linewidth}
        \centering
        \includegraphics[width=0.95\textwidth]{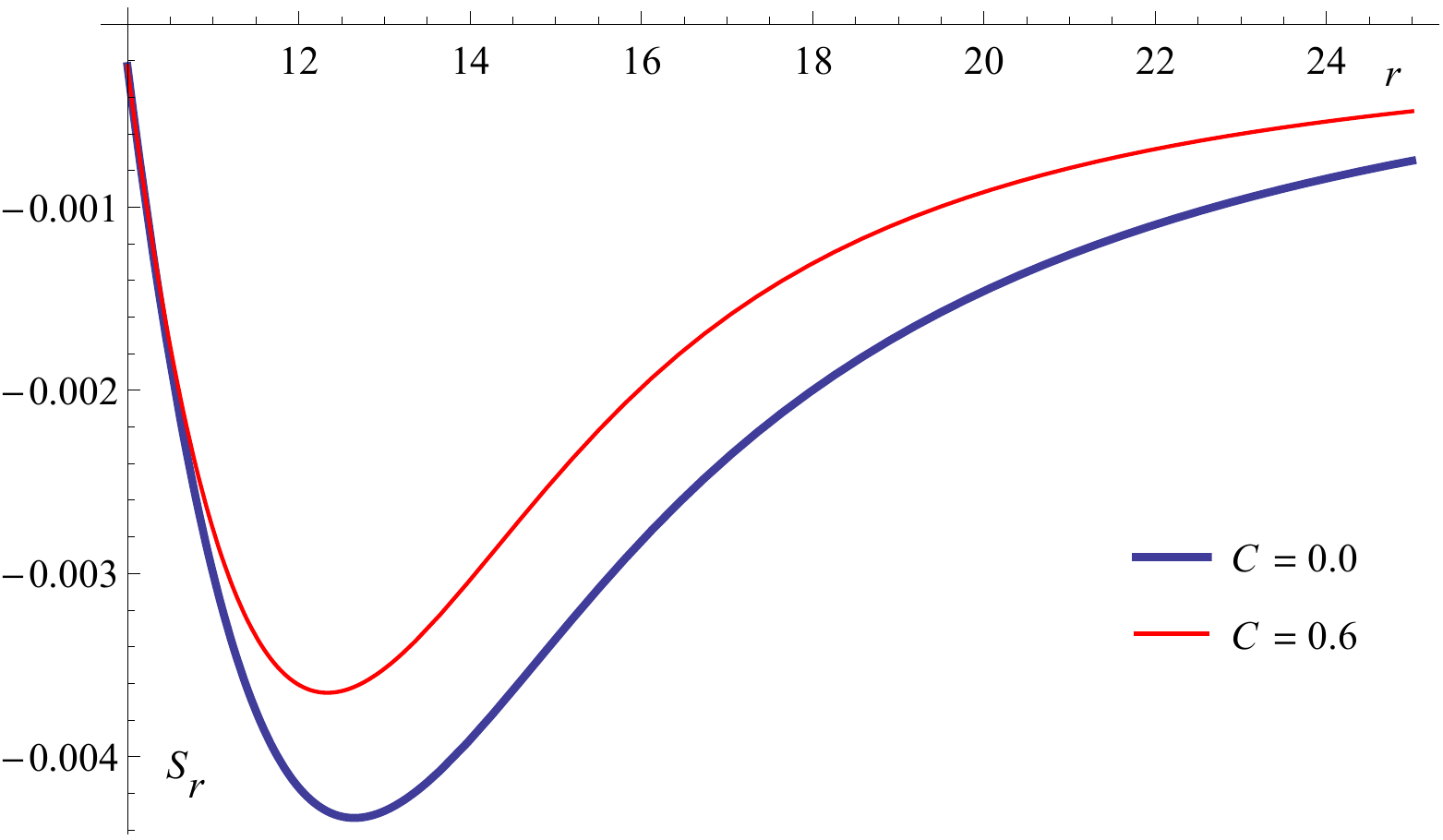}
        \label{fig:figura1minipg}
    \end{minipage}\hfill
    \begin{minipage}{0.5\linewidth}
        \centering
        \includegraphics[width=0.95\textwidth]{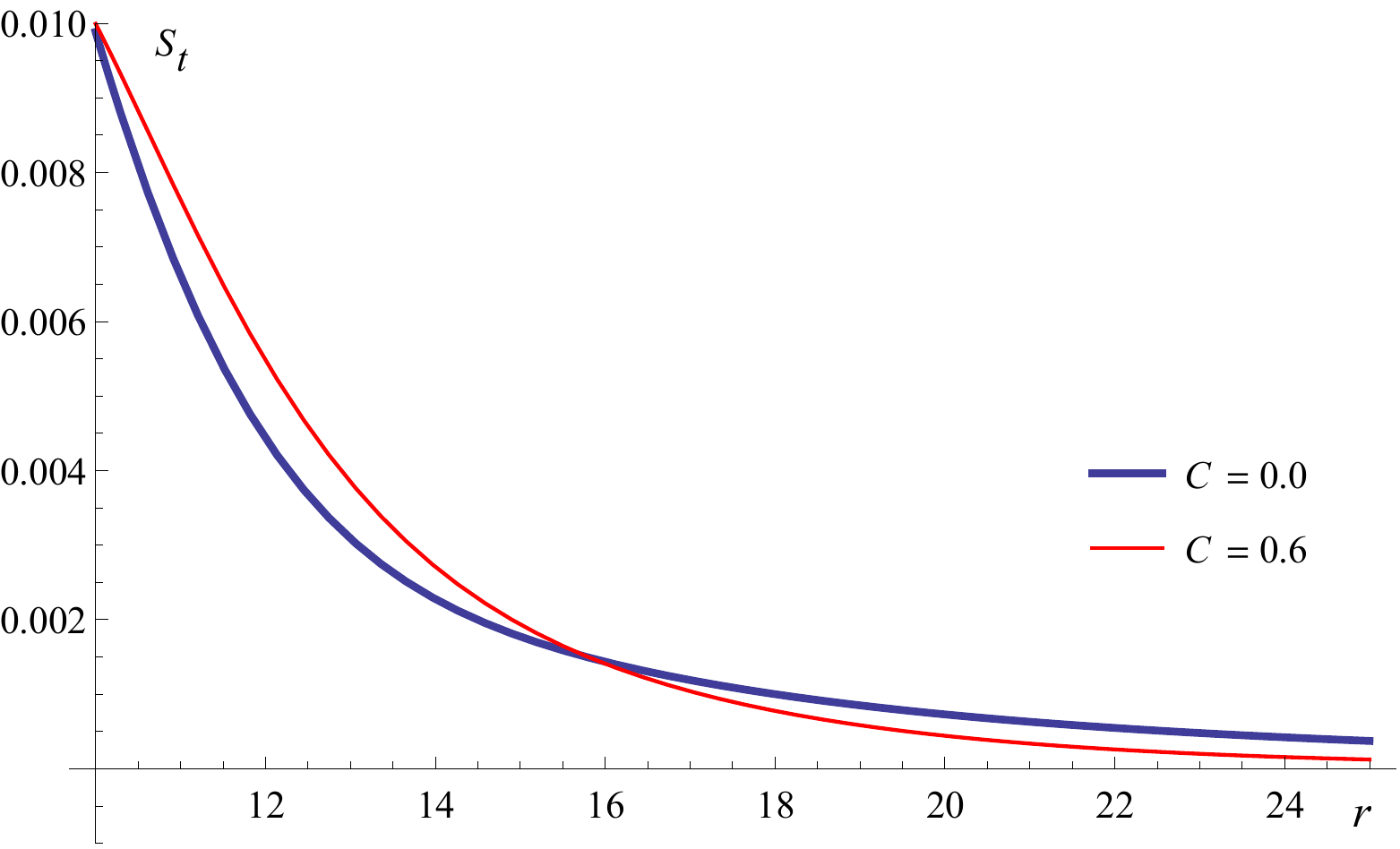}
              \label{fig:figura2minipg}
    \end{minipage}
   \caption{Sum of density with radial (transversal) pressure, in left (right) panel, as a function of the radial coordinate, $r$, for D-S GUP corrected wormholes, considering both zero tidal ($C=0$) and non-zero tidal cases. The parameter settings are $a = 8.0$, $r_0 = 10.0$, $\alpha=1.0$, in Planck units.}
    \label{Energy2}
\end{figure}

Null energy conditions (NEC) are not obeyed by wormholes, at least in the context of GR, provided the flaring-out conditions are valid. Thus, Fig. \ref{Energy2} shows us the behaviors of $S_r=\rho+p_r$ (density with radial pressure - left panel) and $S_t=\rho+p_t$ (density with transversal pressure - right panel), for Dymnikova-Schwinger GUP corrected wormholes. Considering the behavior of $S_r$, the violation is more pronounced for the zero-tidal wormhole ($C=0$).

\begin{figure}[!h]
    \centering
        \includegraphics[width=0.55\textwidth]{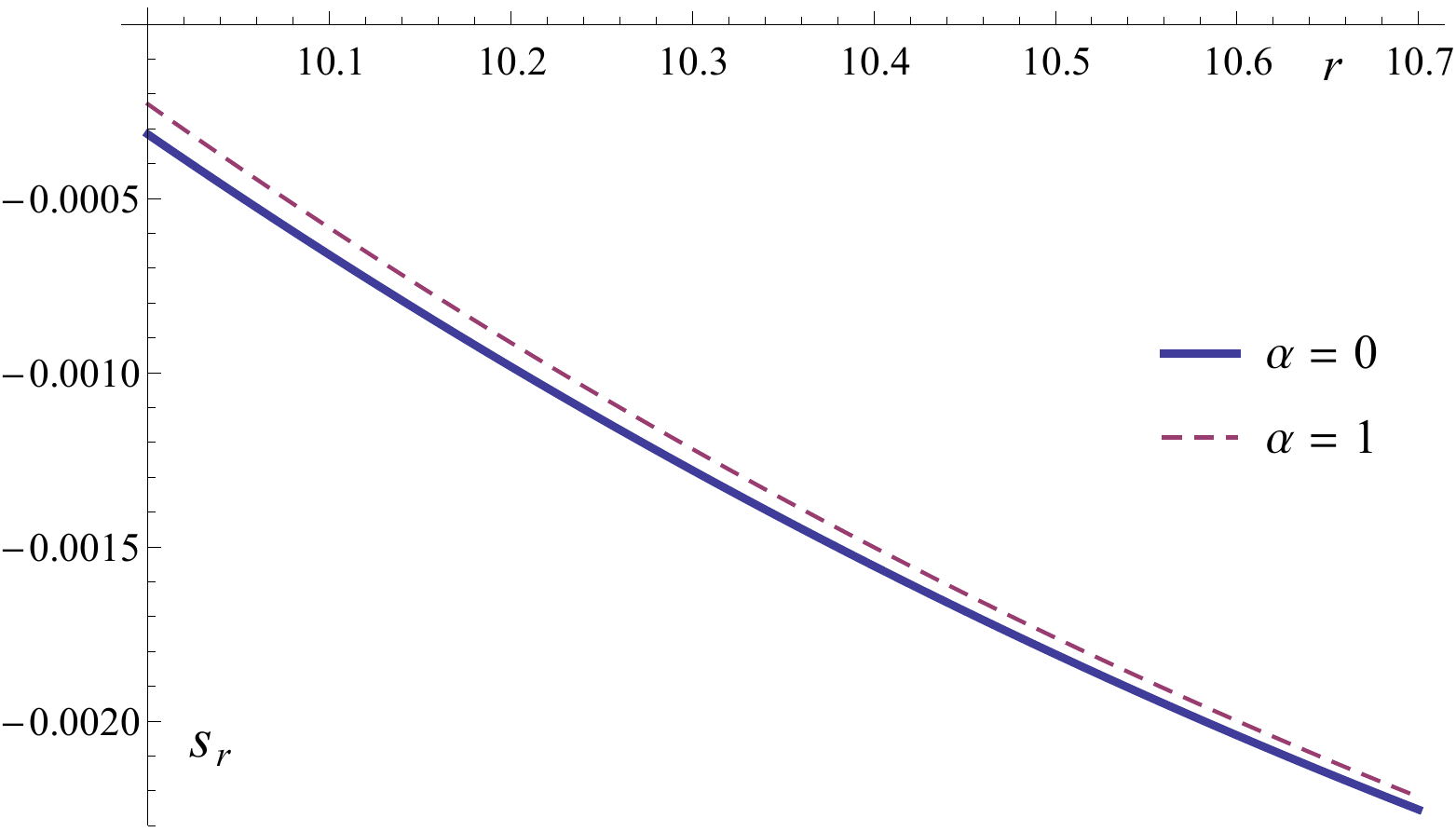}
        \label{fig:figura1minipg}
    \caption{Sum of density with the radial pressure, $S_r$, as a function of the radial coordinate, for $a=8.0$, $r_0=10$, and $C=0.6$, for both Dymnikova-Schwinger ($\alpha=0.0$) and D-S GUP corrected ($\alpha=1.0$) wormholes.} %In the right panel, the parameter state of D-S ($\alpha=0.0$) and D-S GUP corrected ($\alpha=0.1$) wormholes, with parameter settings $a = 0.8$, $r_0 = 1.0$, and $C=0.6$, in Planck units.}
    \label{omega2}
\end{figure}
Now let us analyze the influence of the minimal length on NECs. Although these conditions remain still unfulfilled demanding the presence of exotic matter, the occurrence of a fundamental minimal length attenuates this violation. The left panel of Fig. \ref{omega2} reveals this feature, on exhibiting that $S_r^{\alpha=0}<S_r^{\alpha\neq 0}$, nearby the wormhole throat. 

We can also analyze the exoticity of the material source by considering the state parameter of a perfect fluid in the neighboring of the wormhole throat, according to the linear EoS $\omega(r)=p_r/\rho$. Thus, we have that
\begin{equation}\label{Limit2}
\lim_{r\to r_0}\omega(r)=-\frac{a^3 \left[\exp{\left(\frac{r_0^3}{a^3}\right)}-1\right]}{3 \left(a \alpha +r_0^3\right)}.
\end{equation}
\begin{figure}[!h]
    \centering
        \includegraphics[width=0.55\textwidth]{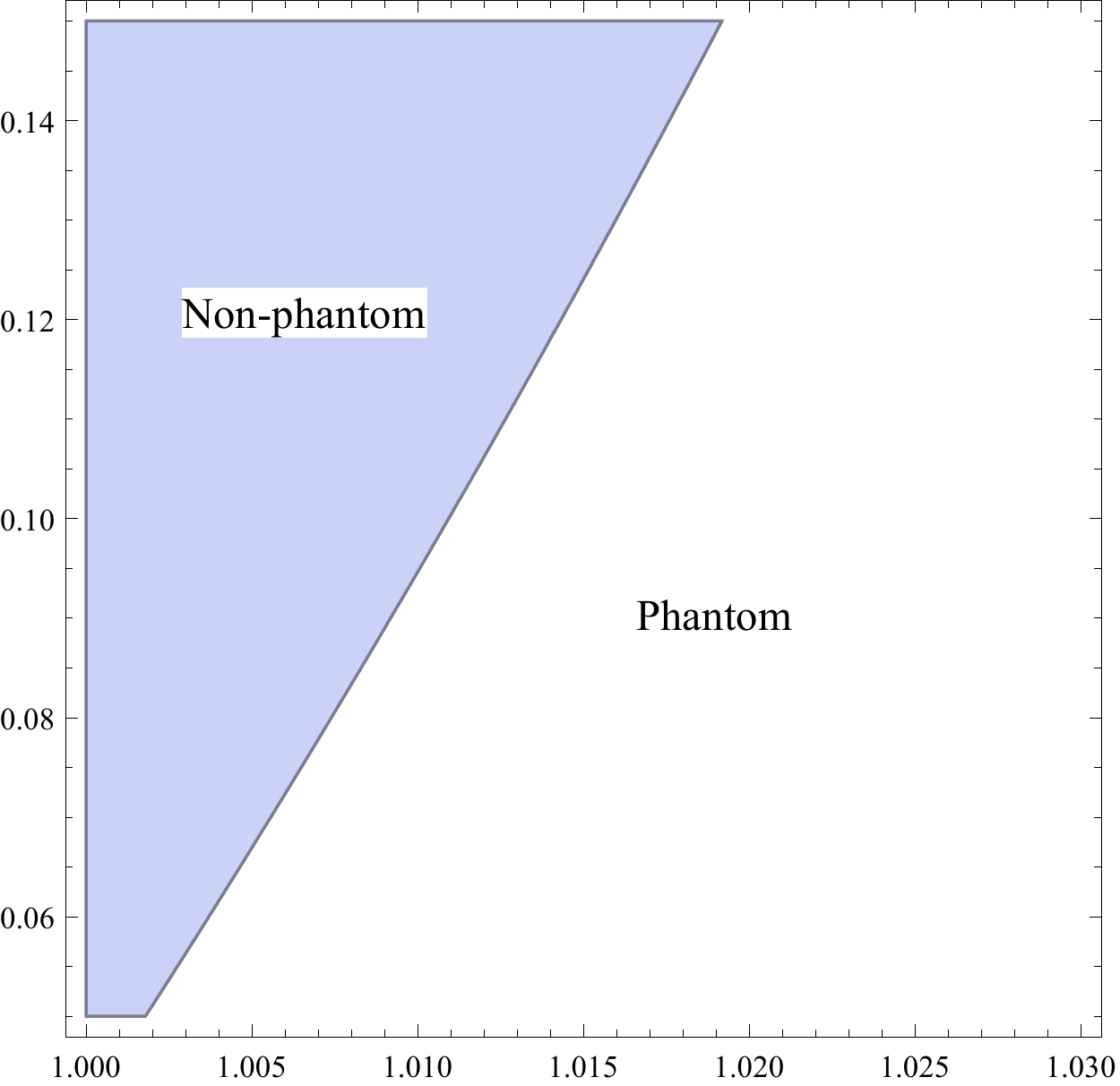}
        \label{fig:figura1minipg}
    \caption{Parameter space ($r_0$, $\alpha$) associated to D-S GUP corrected wormholes, with $a = 0.8$, in Planck units.}
    \label{omega2}
\end{figure}

In Fig. \ref{omega2}, we depict the parameter space ($r_0$,$\alpha$) obtained from Eq. (\ref{Limit2}), remarking the regions where we have a non-phantom ($\omega>-1$) behavior of the source. Notice that this occurs at Planckian scales, and we can interpret such a feature due to the presence of the minimal length, which introduces quantum effects at these scales. This latter explains why the behavior near the wormhole throat differs from the classical GR.

\section{Conclusion}

In this work, we have found new $d$-dimensional and asymptotically flat wormhole solutions with the presence of specific matter fields in the energy-momentum tensor. The energy density and the pressure components have local values for different values of the radial coordinate, and in section \ref{nuestrasCondiciones} we have proposed a list of constraints that the energy density as well as both the redshift and shape functions must satisfy in such a manner that the new wormhole solutions fulfill the corresponding criteria.

Following, we have initially used the $d$-dimensional generalization \cite{Estrada:2019qsu} of the Dymnikova energy density \cite{Dymnikova:1992ux}, and by solving the time component of Einstein's equation we have found the wormhole exact solutions in $d$ spacetime dimensions, which depend on the throat radius, $r_0$, and on a characteristic length, $a$. This latter is a scale associated to the material source.  We have also defined a suited logarithmic redshift function dependent on these parameters and a factor that, on vanishing, one obtains a zero-tidal simplest traversable wormhole. Remarkably the energy density, the redshift function, and the pressure components satisfy the criteria for the building of a wormhole. 

The detailed study of the conditions for the existence of these objects (as the flaring-out one) has shown that there has to be a relationship between $r_0$, $a$, and the spacetime dimension $d$ in order to guarantee their existence. Following, we have generated the embedding diagram for some of these wormholes, showing that on increasing the spacetime dimension the asymptotic flatness is more quickly reached. In other words, the greater $d$ the smaller the slope towards the wormhole throat.

With respect still to $d$-dimensional Dymnikova wormholes, we have analyzed Weak and Null Energy Conditions (WEC and NEC) concerning the material source. As expected, this latter is considered exotic, since the radial part of those conditions is violated -- while the transversal one is not -- and such a violation is more pronounced the larger the spacetime dimension, in regions nearby the wormhole throat. This feature is corroborated by the behavior of the position-dependent state parameter obeying the linear EoS $\omega(r)=p_r/\rho$ around the throat. Its behavior indicates that such a fluid is phantom-like since $\omega(r)<-1$ there.

In sequence, we specialized the previous study for 4d wormholes, which we have called Dymnikova-Schwinger since the material source under consideration is the gravitational analogue of the pair density produced in a vacuum by means of the application of an intense (electric) field -- the so-called Schwinger effect, predicted by QED. The gravitational counterpart of this phenomenon holds some similarity with that one associated to the Casimir effect, which has been widely studied in the context of wormholes \cite{Garattini:2019ivd,Jusufi:2020rpw,Tripathy:2020ehi,Carvalho:2021ajy,Oliveira:2021ypz}. Thus, on considering the correction to the Schwinger effect due to the introduction of a minimal length ($\ell=\sqrt{\alpha}$) via GUP \cite{Haouat:2013yba,Ong:2020tvo}, we have found the corresponding correction to the gravitational analogue of Dymnikova's matter distribution. Supposing that the minimal length is very small, we obtained a novel traversable and asymptotically flat wormhole solution that reduces to the one previously studied when $\alpha=0$ and $d=4$, for the same redshift function employed before.

The corresponding embedding diagram shows that the presence of the minimal length increases the slope of the wormhole towards its throat compared with the case without such a length -- in other words, it lessens the wormhole flatness. On the other hand, the WEC and NEC analysis have shown that this quantity, although still yields the non-fulfillment of those conditions, attenuates their violation since $\rho+p_r$ increases with it. Finally, the study of the linear state parameter $\omega(r)$ has revealed that, in the presence of the fundamental length, the source ceases to be a phantom-type nearby the wormhole throat, at least at Planck's scale. This behavior should be attributed to quantum effects coming from GUP, revealing why it differs from that predicted by classical GR.

\acknowledgments Celio R. Muniz thanks the Conselho Nacional de Desenvolvimento Cient\'{i}fico e Tecnol\'{o}ogico (CNPq), grant no 308268/2021-6 for financial support. Milko Estrada is funded by ANID , FONDECYT de Iniciaci\'on en Investigación 2023, Folio 11230247.


\newpage

\begin{thebibliography}{99}



    \bibitem{Morris:1988cz}
  M.~S.~Morris and K.~S.~Thorne,
  5``Wormholes in space-time and their use for interstellar travel:
  A tool for teaching general relativity,''
  Am.\ J.\ Phys.\  {\bf 56}, 395 (1988).
  %%CITATION = AJPIA,56,395;%%

  \bibitem{Visser}
  M.~Visser,
 ``Lorentzian Wormholes: From Einstein to Hawking'',
   American Institute of Physics (1996).

 \bibitem{Maldacena:2013xja}
 J.~Maldacena and L.~Susskind,
%``Cool horizons for entangled black holes'',
Fortsch. Phys. {\bf 61}, 781 (2013),
 doi: 10.1002/prop.201300020,
arxiv: 1306.0533 [hep-th].

\bibitem{Maldacena:2017axo}
J.~Maldacena, D.~Stanford, and Z.~Yang,
%``Diving into traversable wormholes'',
Fortsch. Phys., {\bf 65} 1700034 (2017).
   doi: 10.1002/prop.201700034,
arXiv: 1704.05333 [hep-th].

\bibitem{Maldacena:2020sxe}
J.~Maldacena and A.~Milekhin,
%``Humanly traversable wormholes'',
Phys. Rev. {\bf D103}, 066007 (2021),
doi: 10.1103/PhysRevD.103.066007,
arXiv: 2008.06618 [hep-th].

\bibitem{Gonzalez:2009je}
J.~Gonzalez and J.~Herrero,
%``Graphene wormholes: A Condensed matter illustration of Dirac fermions in curved space'',
Nucl. Phys. {\bf B825}, 426 (2010),
 doi: 10.1016/j.nuclphysb.2009.09.028,
arXiv: 0909.3057 [cond-mat.mes-hall].

\bibitem{Alencar:2021ejd}
G.~Alencar, V.~B.~Bezerra, and C.~R.~Muniz,
%``Casimir wormholes in $2+1$ dimensions with applications to the graphene'',
Eur. Phys. J. {\bf C81}, 924 (2021),
doi: 10.1140/epjc/s10052-021-09734-0,
arXiv: 2104.13952 [gr-qc].

\bibitem{Pavlovic:2014gba}
P.~Pavlovic and M.~Sossich,
%``Wormholes in viable $f(R)$ modified theories of gravity and Weak Energy Condition'',
Eur. Phys. J. {\bf C75}, 117 (2015),
doi: 10.1140/epjc/s10052-015-3331-y,
arXiv: 1406.2509 [gr-qc].

\bibitem{Myrzakulov:2015kda}
R.~Myrzakulov, L.~Sebastiani, S.~Vagnozzi, and S.~Zerbini, 
%    title = "{Static spherically symmetric solutions in mimetic gravity: rotation curves and wormholes}",
Class. Quant. Grav., {\bf33}, 12, 125005 (2016),
doi: 10.1088/0264-9381/33/12/125005
arXiv: 1510.02284, [gr-qc].
         
\bibitem{Mehdizadeh:2019qvc}
M.~R.~Mehdizadeh and A.~H.~Ziaie,
%``Traversable wormholes in Einsteinian cubic gravity'',
Mod. Phys. Lett. {\bf A35}, 2050017 (2019),
doi: 10.1142/S0217732320500170,
arXiv: 1903.10907 [gr-qc].

 \bibitem{Sahoo:2020sva}
 P.~Sahoo, P.~H.~R.~S.~Moraes, M.~M.~Lapola, and P.~K.~Sahoo,
%``Traversable wormholes in the traceless f(R,T) gravity'',
Int. J. Mod. Phys. {\bf D30}, 2150100 (2021).
 doi: 10.1142/S0218271821501005,
 arXiv: 2012.00258, [gr-qc].

 \bibitem{Moti}
 R.~Moti and A.~Shojai,
%`` Traversability of quantum improved wormhole solution'',
 Phys. Rev. {\bf D101}, 124042 (2020),
doi:10.1103/PhysRevD.101.124042
arXiv: 2006.06190 [gr-qc].

\bibitem{Alencar}
G.~Alencar, V.~B.~Bezerra, C.~R.~Muniz, and H.~S.~Vieira, Universe 7, 238 (2021),
%``Ellis–Bronnikov Wormholes in Asymptotically Safe Gravity'',
doi:10.3390/universe7070238
arXiv:2106.02476 [gr-qc].

\bibitem{Sadeghi:2022sto}
J.~Sadeghi, B.~Pourhassan, S.~N.~ Gashti, and S.~Upadhyay,
%``Smeared mass source wormholes in modified f(R) gravity with the Lorentzian density distribution function'',
Mod. Phys. Lett. {\bf A37}, 2250018 (2022),
 doi: 10.1142/S0217732322500183,
arXiv: 2203.04543 [gr-qc].

\bibitem{Nilton:2022cho}
M.~Nilton, J.~Furtado, G.~Alencar,
%``Traversability of wormhole solutions in asymptotically safe gravity'',
Phys. Rev. {\bf D105}, 084048 (2022),
doi:10.1103/PhysRevD.105.084048,
arXiv: 2202.04188 [gr-qc].

\bibitem{Harko:2013yb}
T.~Harko, F.~S.~N.~Lobo, M.~K.~Mak and S.~V.~Sushkov,
%``Modified-gravity wormholes without exotic matter,''
Phys. Rev. D \textbf{87}, no.6, 067504 (2013)
doi:10.1103/PhysRevD.87.067504
[arXiv:1301.6878 [gr-qc]].

\bibitem{Bronnikov:2016xvj}
K.~A.~Bronnikov and A.~M.~Galiakhmetov,
%``Wormholes and black universes without phantom fields in Einstein-Cartan theory,''
Phys. Rev. D \textbf{94}, no.12, 124006 (2016)
doi:10.1103/PhysRevD.94.124006
[arXiv:1607.07791 [gr-qc]].

\bibitem{Kuhfittig:2020zmp}
P.~K.~F.~Kuhfittig,
%``Noncommutative-geometry wormholes without exotic matter,''
Adv. Stud. Theor. Phys. \textbf{14}, no.5-8, 219-225 (2020)
[arXiv:2008.06728 [gr-qc]].

\bibitem{Chanda:2021dvc}
A.~Chanda, S.~Dey and B.~C.~Paul,
%``Morris\textendash{}Thorne wormholes in modified $f(R,T)$ gravity,''
Gen. Rel. Grav. \textbf{53}, no.8, 78 (2021)
doi:10.1007/s10714-021-02847-7
[arXiv:2102.01556 [gr-qc]]

\bibitem{Paul:2021lvb}
B.~C.~Paul,
%``Emergent universe in $D \ge 4$ dimensions with dynamical wormholes,''
Eur. Phys. J. C \textbf{81}, no.8, 776 (2021)
doi:10.1140/epjc/s10052-021-09562-2

\bibitem{Oliveira:2021ypz}
P.~H.~F.~Oliveira, G.~Alencar, I.~C.~Jardim and R.~R.~Landim,
%``Traversable Casimir wormholes in D dimensions,''
Mod. Phys. Lett. A \textbf{37}, no.15, 2250090 (2022)
doi:10.1142/S0217732322500900
[arXiv:2107.00605 [hep-th]]

\bibitem{Garattini:2019ivd}
R.~Garattini,
%    title = "{Casimir Wormholes}",
Eur. Phys. J. {\bf C79}, 11, 951 (2019),
doi: 10.1140/epjc/s10052-019-7468-y,
[arXiv: 1907.03623 [gr-qc]].

\bibitem{Jusufi:2020rpw}
K.~Jusufi, P.~Channuie, and M.~Jamil,
 %   title = "{Traversable Wormholes Supported by GUP Corrected Casimir Energy}",
 Eur. Phys. J. {\bf C80}, 2, 127 (2020),
 doi: 10.1140/epjc/s10052-020-7690-7,
[arXiv: 2002.01341 [gr-qc]].

\bibitem{Tripathy:2020ehi}
S.~K.~Tripathy, 
%    title = "{Modelling Casimir wormholes in extended gravity}",
Phys. Dark Univ. {\bf31}, 100757 (2021),
 doi: 10.1016/j.dark.2020.100757,
[arXiv: 2004.14801 [gr-qc]].

\bibitem{Carvalho:2021ajy}
I.~D.~.D.~Carvalho, G.~Alencar, and C.~R.~Muniz,
 %   title = "{Gravitational bending angle with finite distances by Casimir wormholes}",
 Int. J. Mod. Phys. {\bf D31}, 03, 2250011 (2022),
  doi: 10.1142/S0218271822500110
[arXiv: 2106.11801 [gr-qc]].

\bibitem{Cataldo:2015vra}
M.~Cataldo, L.~Liempi and P.~Rodr\'\i{}guez,
%``Morris-Thorne wormholes in static pseudo-spherically symmetric spacetimes,''
Phys. Rev. D \textbf{91}, no.12, 124039 (2015)
doi:10.1103/PhysRevD.91.124039
[arXiv:1506.04685 [gr-qc]].

\bibitem{Tello-Ortiz:2021kxg}
F.~Tello-Ortiz, S.~K.~Maurya and P.~Bargue\~no,
%``Minimally deformed wormholes,''
Eur. Phys. J. C \textbf{81}, no.5, 426 (2021)
[erratum: Eur. Phys. J. C \textbf{82}, no.8, 742 (2022)]
doi:10.1140/epjc/s10052-021-09179-5

\bibitem{Tello-Ortiz:2020zfs}
F.~Tello-Ortiz and E.~Contreras,
%``Traversable wormholes in light of class I approach,''
Annals Phys. \textbf{419}, 168217 (2020)
doi:10.1016/j.aop.2020.168217

\bibitem{Xu:2020wfm}
Z.~Xu, M.~Tang, G.~Cao and S.~N.~Zhang,
%``Possibility of traversable wormhole formation in the dark matter halo with istropic pressure,''
Eur. Phys. J. C \textbf{80}, no.1, 70 (2020)
doi:10.1140/epjc/s10052-020-7636-0

\bibitem{Muniz:2022eex},
C.~R.~Muniz and R.~V.~Maluf,
%    title = "{A class of traversable wormholes in the Starobinsky-like f(R) gravity with anisotropic dark matter}",
Annals Phys. {\bf446}, 169129 (2022),
doi: 10.1016/j.aop.2022.169129.

\bibitem{Garattini:2021kca}
R.~Garattini,
%``Yukawa\textendash{}Casimir wormholes,''
Eur. Phys. J. C \textbf{81}, no.9, 824 (2021)
doi:10.1140/epjc/s10052-021-09634-3
[arXiv:2107.09276 [gr-qc]]

\bibitem{Estrada:2019qsu}
M.~Estrada and R.~Aros,
%``Regular black holes with $\Lambda>0$ and its evolution in Lovelock gravity,''
Eur. Phys. J. C \textbf{79}, no.10, 810 (2019)
doi:10.1140/epjc/s10052-019-7316-0
[arXiv:1906.01152 [gr-qc]].

\bibitem{Dymnikova:1992ux}
I.~Dymnikova,
%``Vacuum nonsingular black hole,''
Gen. Rel. Grav. \textbf{24}, 235-242 (1992)
doi:10.1007/BF00760226.

\bibitem{Dymnikova:1996plb}
I.~G.~Dymnikova,
%``De Sitter-Schwarzschild black hole: Its particlelike core and thermodynamical properties'',
Int.J.Mod.Phys. {\bf D5}, 5, 529 (1996),
doi: 10.1142/s0218271896000333.



\bibitem{Ansoldi:2008jw}
S.~Ansoldi,
%``Spherical black holes with regular center: A Review of existing models including a recent realization with Gaussian sources''
Conference on Black Holes and Naked Singularities,
reportNumber: KUNS-2108, 2 (2008),
[arXiv: 0802.0330 [gr-qc]].


\bibitem{Haouat:2013yba}
S.~Haouat and K.~Nouicer,
%``Influence of a Minimal Length on the Creation of Scalar Particles'',
Phys. Rev. {\bf D89}, 10, 105030 (2014),
 doi: 10.1103/PhysRevD.89.105030,
[arXiv: 1310.6966 [hep-th]].


\bibitem{Ong:2020tvo}
Y.~C.~Ong,
%``Schwinger pair production and the extended uncertainty principle: can heuristic derivations be trusted?''
Eur. Phys. J. {\bf C80}, 8, 777 (2020),
    doi: 10.1140/epjc/s10052-020-8363-2,
    [arXiv: 2005.12075 [gr-qc]].


\bibitem{Aros:2019quj}
R.~Aros and M.~Estrada,
%``Regular black holes and its thermodynamics in Lovelock gravity,''
Eur. Phys. J. C \textbf{79}, no.3, 259 (2019)
doi:10.1140/epjc/s10052-019-6783-7
[arXiv:1901.08724 [gr-qc]].

\bibitem{Spallucci:2017aod}
E.~Spallucci and A.~Smailagic,
%``Regular black holes from semi-classical down to Planckian size,''
Int. J. Mod. Phys. D \textbf{26}, no.07, 1730013 (2017)
doi:10.1142/S0218271817300130
[arXiv:1701.04592 [hep-th]].




\bibitem{Romero:2019ull}
J.~M.~Romero and M.~Bellini,
%``Traversable wormhole magnetic monopoles from Dymnikova Metric,''
Eur. Phys. J. Plus \textbf{134}, no.11, 579 (2019)
doi:10.1140/epjp/i2019-12926-1
[arXiv:1906.00062 [gr-qc]].

\end{thebibliography}
 \end{document}